\documentclass[%
 aip,
 amsmath,amssymb,
 reprint,%
]{revtex4-1}

\usepackage{graphicx}
\usepackage{dcolumn}
\usepackage{bm}
\usepackage{xcolor} 

\usepackage[utf8]{inputenc}
\usepackage[T1]{fontenc}
\usepackage{mathptmx}

\begin{document}

\preprint{AIP/123-QED}

\title{Magnetic correlations in the disordered ferromagnetic alloy Ni-V revealed with small angle neutron scattering}

\author{A. Schroeder}
 \email{aschroe2@kent.edu}
 \author{S. Bhattarai}
\author{A Gebretsadik}
\altaffiliation[Present address:] { Intel, Chandler AZ, USA}
\author{H. Adawi}
\author{J.-G. Lussier}
 \affiliation{ Department of Physics, Kent State University, Kent OH 44242, USA}%

\author{K. L. Krycka}
\affiliation{National Institute of Standards and Technologies, NIST Center of Neutron Research, Gaithersburg MD 20899, USA}

\date{\today}

\begin{abstract}
We present small angle neutron scattering (SANS) data collected on polycrystalline Ni$_{1-x}$V$_x$ samples with $x\geq0.10$ with confirmed random atomic distribution. We aim to determine the relevant length scales of magnetic correlations in  ferromagnetic samples with low critical temperatures $T_c$ that show signs of magnetic inhomogeneities in magnetization and $\mu$SR data. The SANS study reveals signatures of long-range order and coexistence of short-range magnetic correlations in this randomly disordered ferromagnetic alloy. We show the advantages of a polarization analysis  in identifying the main magnetic contributions from the dominating nuclear scattering. 
 
\end{abstract}

\maketitle


Formation of ferromagnetism in metals is still an active field for discovery of novel phases and mechanisms in condensed matter physics\cite{Brando2016}. In particular, the control of disorder and determination of how inhomogeneities 
affect magnetic properties remains  a significant challenge. 
Small angle neutron scattering (SANS) is one of the prime methods\cite{Muhlbauer2019} to characterize magnetic material at the nanoscale. 
 It has revealed important insight in the complex structure formation
 of inhomogeneous magnets with defects or internal structures from bulk alloys\cite{Verbeek1980} to amorphous and nanocrystalline magnetic materials\cite{Muhlbauer2019,Michels2008}.
In this study we focus on the binary transition metal alloy Ni$_{1-x}$V$_x$, that presents an example of a diluted inhomogeneous ferromagnet produced by random atomic distribution\cite{Wang2017,AG}.  The onset of ferromagnetic order of Ni at { $T_c=630$~K} is suppressed towards zero with sufficient V concentration of {$x_c=0.116$\,\cite{PRL}}. Previous magnetization and $\mu$SR studies show signatures of fluctuating clusters  \cite{PRL} from Ni-rich regions for paramagnetic samples with $x> x_c$. These persist also into the ferromagnetic state close to $x_c$ and coexist with the static order \cite{Wang2017} evolving below $T_c$. With SANS we aim to measure the magnetic cluster sizes and their effect on the static order  
in this random disordered system. 
We present a SANS study with polarization analysis \cite{Krycka2012} to extract magnetic scattering that would otherwise be  dominated by nuclear scattering.\\


For this study we used the same polycrystalline samples of Ni$_{1-x}$V$_x$ that were prepared for optimal random distribution and characterized by several methods \cite{AG} from previous studies\cite{PRL,Wang2017}. 
Several pellets of {3\,mm} diameter of each concentration were wrapped in Al foil and mounted on Al- sample holder framed with Cd-mask and connected to the cold plate of the cryostat.
The SANS experiments were performed at GPSANS, HFIR, Oak Ridge National Lab and at NG7SANS\cite{Glinka1998}, NCNR, NIST.
We show detailed data from NIST of Ni$_{0.90}$V$_{0.10}$ samples using also polarized neutrons (tracking the polarization (p) state before and after sample). The SANS intensity was collected in the $xy$-plane on a 2D detector at different distances to cover a wave vector ($Q$)-range of {(0.06-1)~nm$^{-1}$} with neutron wavelengths of {0.55\,nm and 0.75\,nm}. 
Taking advantage of supermirror polarizer and $^3$He-cell as spin analyser as described in detail before\cite{Krycka2009,Chen2009} we collected separately non spin flip (NSF) scattering with unchanged p-state of the neutrons (DD and UU) and spin flip (SF) scattering with reversed p-state (DU and UD) from the sample. U, D refers to the neutron spins aligned UP, DOWN with respect to the neutron polarization axis defined by the external magnetic field. 
The magnetic field was applied in the $x$-direction {($B_{max}=1.5$\,T, $B_{min}=7$\,mT)} perpendicular to the beam ($\parallel z$). $\theta$ indicates the azimuthal angle within the $xy$-plane, with {$\theta=0^{\circ}$} in the horizontal $x$-direction. Horizontal or vertical data averages included a symmetric cone of width $\pm \Delta \theta =30^\circ$ around $\theta=0^\circ$ and $180^\circ$ or $\theta=\pm90^\circ$ presented as SFH or SFV, respectively.  
Besides fully polarization-analyzed  PASANS we collected also unpolarized or "non pol" data (NP) without tracking the p-state and "half pol" (HP) data without distinguishing the p-state after the sample (D and U) without a $^3$He-cell in the beam. The PASANS data were polarization corrected and reduced with the IGOR software\cite{Kline2006}. \\     

%
%

First SANS studies of Ni$_{1-x}$V$_x$ could not resolve the weak magnetic scattering for paramagnetic samples with $x=0.12$ but recent SANS data collected at NG7SANS, NIST and GPSANS, ORNL revealed a clear temperature-dependent signal  in weak ferromagnetic samples with $x\leq0.11$ \cite{Shiva} that demonstrated that magnetic scattering can be resolved for ferromagnetic samples with a reduced average moment per Ni of {$\mu \approx0.03 \;\mu_B$}\cite{Wang2017}. We show here the SANS data for $x=0.10$ with {$T_c\approx50$~K} collected at NIST to compare best full pol   and non pol data. Fig.\,1(a) presents the non polarized neutron scattering intensity as a function of the magnitude of the wavevector $Q$ collected as horizontal average $NPH$.
$NPH$ collects  only transverse magnetic contributions $M_y^2$ and $M_z^2$ (not the longitudinal magnetic components $M_x^2$ along the field direction) beside the dominating nuclear contribution $N^2$ and other backgrounds $BG_{NP}$.  
\begin{equation}
NPH=NP(\theta=0^{\circ})=N^2+M_z^2+M_y^2+BG_{NP}
\end{equation}
%
The ``non-magnetic'' contributions are estimated by $NPH$ collected in high fields {($NPH(B=1.5$\,T, $T<T_c$))}
where aligned magnetic moments are expected to contribute only to the magnetic scattering in field direction $M_x^2$ not to the selected transverse components.   
The $NPH$ difference of the total non pol scattering collected at different temperatures in low fields {($B=7$~mT)} and the high field data should finally reveal the {\it magnetic scattering} ($M_z^2+M_y^2$) as shown in Fig.\,1(a). 

Most magnetic scattering is found close to {$T_c\approx 50$~K} in the higher {$Q$ range} {(0.2\,nm$^{-1}$ - 1\,nm$^{-1}$)}. It can be approximated by a Lorentzian as expected for paramagnetic critical scattering of the Ornstein Zernike form with a correlation length $1/\kappa$ increasing towards $T_c$ . Below $T_c$ the magnetic intensity is significantly reduced in the higher {$Q$ regime}, and in addition, an increase of intensity at low $Q$ is noticed that follows a  $1/Q^4$ dependence without any sign of saturation. 
\begin{equation}
I_{mag}(Q)=\frac{D}{Q^4}+\frac{L}{\kappa^2+Q^2} \, \times F(Q)
\end{equation}  
The fit can be improved somewhat towards higher $Q$ by a {form factor\cite{Saurel10}}  $F(Q)=exp(-\frac{1}{5}r_0^2 Q^2)$ with radius {$r_0\approx1$~nm} that could indicate non-uniform magnetic scattering centers or be an artifact of non ideal background subtraction.    
Different than in homogeneous systems with a narrow critical regime with diverging correlations at $T_c$ typically observed with SANS\cite{Lynn1998} we find 
reduced length scales as seen in inhomogeneous ferromagnets\cite{Muhlbauer2019} e.g. in diluted ferromagnetic alloys\cite{Verbeek1980,Burke1982} or diluted manganites\cite{DeTeresa2006}. 
 The correlation length of the visible magnetic contribution remains finite at $T_c$ {($1/\kappa \approx 5$~nm)} and even seems to grow further below $T_c$  at 30\,K.  
At very low {$T=3$~K} we still recognize similar transverse magnetic contributions as observed for $T_c$ with reduced amplitude but similar correlation length. This indicates that short-range fluctuations of the paramagnetic state are still left in the ferromagnetic state at low temperatures. 
The {low-$Q$ upturn} in the non pol data ($NPH$) is most likely due to the contrast of misaligned magnetic regions (domains) of a large scale {($>100$~nm)} expected in a soft magnet of finite size at small fields.
This domain term becomes apparent  below $T_c$ indicating the onset of long-range order. 
However, it is difficult to extract the magnetic response from the huge nuclear $1/Q^4$ term due to grain boundaries in these polycrystalline samples. 
%
%
\begin{figure}
\includegraphics{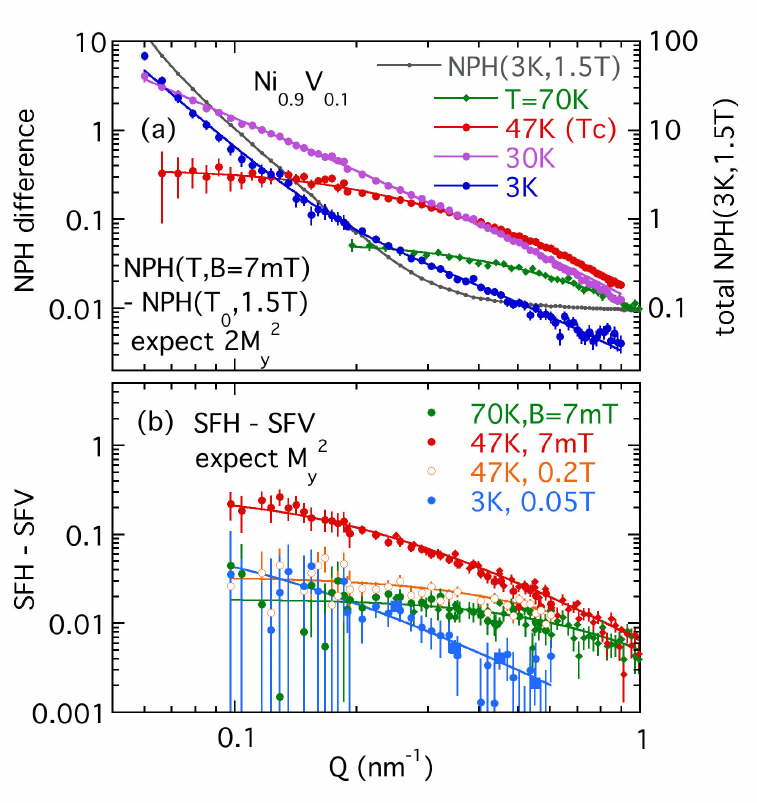}
\caption{\label{fig:both} Magnetic neutron scattering intensity of Ni$_{0.90}$V$_{0.10}$ vs wave vector $Q$. (a) shows non polarized horizontal averaged data {$NPH$} collected at different temperatures in low fields {(7\,mT)} after BG subtraction of high field data, shown separately on a different scale. (b) shows polarized intensity, the spin flip contrast SFH - SFV for different temperatures and magnetic fields indicated. Solid lines present fits using Eq.\,(2). }
\end{figure}
%
%
%
%
%

Encouraged by these promising findings of sufficient magnetic scattering in the larger {$Q$ regime}\cite{Shiva}, but uncertain about reasonable background estimates, 
we collected full polarized SANS. 
The clear advantage is the collection of pure spin flip (SF) data (DU+UD) recognizing electronic magnetic scattering through the angle  {$\theta$ dependence\cite{Krycka2012}} at constant $Q$.
\begin{equation}
SF(\theta)= M_z^2+M_y^2 cos^4\theta+M_x^2 cos^2\theta sin^2\theta + BG_{SF}    
\end{equation}
We noticed signs of anisotropy only in field direction $M_x>0$ (see below) and did not consider transverse terms, $M_y=M_z=0$ simplifies the spin flip  {intensity\cite{Krycka2012}} SF($\theta$) in Eq.\,(3). 
%
%
%
\begin{figure}
\includegraphics{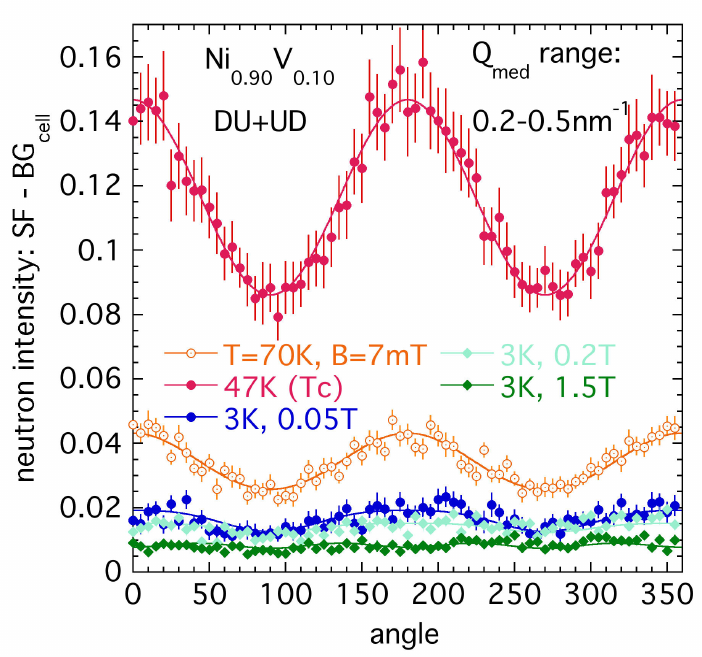}
\caption{\label{fig:SF}Spin flip (SF) intensity of Ni$_{0.90}$V$_{0.10}$ vs angle $\theta$ at constant $Q$ (average over $Q$ range {from 0.2\,nm$^{-1}$ to 0.5\,nm$^{-1}$}) at different temperatures $T$ and magnetic fields $B$ ($\theta=0$). Solid lines are fits using Eq.\,(3). }
\end{figure}
Fig.\,2 presents the {angle dependence} of the SF data collected for a medium $Q$ range, {(0.2 - 0.5)\,nm$^{-1}$}, for Ni$_{0.90}$V$_{0.10}$ after a constant background has been subtracted. 
The solid lines represent fits using Eq.\,(3) that yield $M_y^2$ but also $M_x^2$ with less precision as presented in Fig.\,3(a).  At high temperatures and very small magnetic fields SF($\theta$) follows a pure $cos^2\theta$ dependence with $M_x^2=M_y^2$ expected for isotropic paramagnetic fluctuations. The data confirm also that $M_z^2=M_y^2$. At low {$T=3$~K} the SF data are shown for {50\,mT} and higher fields. 
At smaller fields the ferromagnetic sample (below $T_c$) depolarizes the neutron beam that PASANS cannot be analyzed.    
The magnetic signal $M_y^2$ at {3\,K} in {50\,mT} is reduced to about 10\% of $M_y^2$ at $T_c$. $M_x^2$ is still similar to $M_y^2$ in small fields. In higher fields $M_y^2$ gets suppressed to reach very small values for {1.5\,T}.
The confirmed isotropy underlines the fact that these  {short-range} correlations stem from dynamic fluctuations 
that cold neutrons can collect up to the order of THz especially on smaller ranges.  
These SF data demonstrate that indeed a small fraction of magnetic fluctuations remain at low temperatures, that get further suppressed in higher magnetic fields. 

%
\begin{figure}
\includegraphics{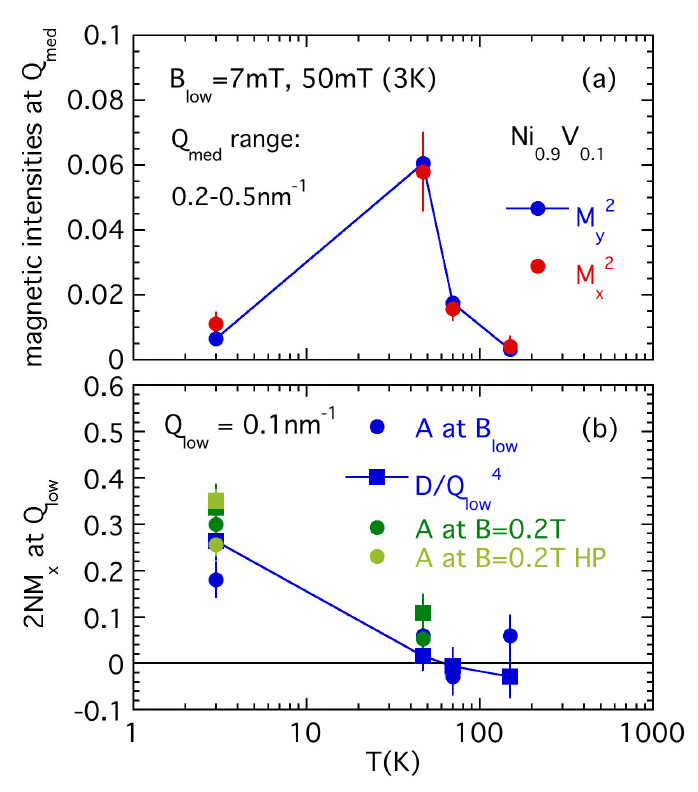}
\caption{\label{fig:Tdep} (a) {Temperature($T$) dependence} of magnetic components $M_y^2$ and $M_x^2$ at  {$Q_{med}\approx0.35\,\mathrm{nm}^{-1}$} from fit of Eq.\,(3) (see Fig.\,2) tracing the evolution of magnetic clusters in Ni$_{0.90}$V$_{0.10}$.  (b) {$T$ dependence} of the interference term $2NM_x$ at  {$Q_{low}\approx0.1\,\mathrm{nm}^{-1}$} indicating the evolution of long-range magnetic domains. $D/Q_{low}$ is evaluated from $DIFV(Q)$  with Eq.\,(2) and $A$ is maximum value in $DIF(\theta)$  using Eq.\,(4) (see Fig.\,4).}
\end{figure}
%
%
%
%
\begin{figure}
\includegraphics{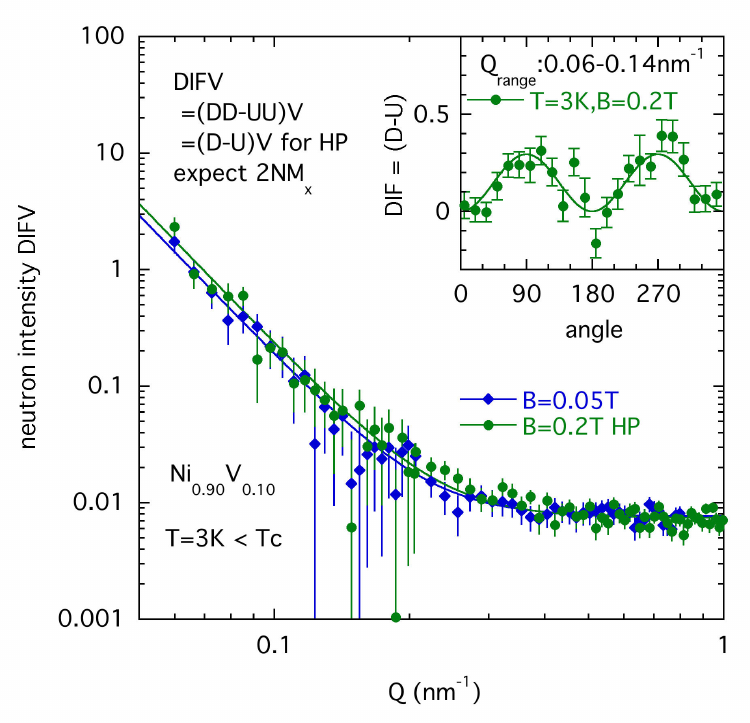}
\caption{\label{fig:NM} $DIF$ or flipper difference from full pol NSF data (DD-UU) and half pol data (D-U) to reveal interference term $2NM_x$ indicating magnetic anisotropy with positive $M_x$ in Ni$_{0.90}$V$_{0.10}$ for low  {$T=3$~K} in magnetic fields of  {0.05\,T and 0.2\,T}. The main panel shows the {$Q$ dependence} of $DIFV$. Solid lines follow Eq.\,(2) with large $\kappa$. The inset presents the {$\theta$ dependence} of $DIF$ at  {$Q_{low}\approx 0.1 \,\mathrm{nm}^{-1}$}. The solid line is a fit that follows Eq.\,(4). }
\end{figure}

Fig.\,1(b) shows the {$Q$ dependence} of the SF contrast, SFH-SFV, that evaluates $M_y^2$. 
Since the magnetic response is isotropic for low fields this signal represents ($1/3$ of) the total magnetic fluctuations $M_{tot}^2$. In this restricted $Q$ regime a Lorentzian describes the data well. A finite {$r_0\approx1$\,nm} also improves the fit somewhat for large $Q$. We did not aim to get more detailed nanostructures from these data using alternative descriptions including cluster distributions\cite{Calderon2005}. The Lorentzian fit produces a correlation lengths for {47\,K} in the order of {7\,nm} that is similar at {$T=3$~K in 50\,mT}. { The estimate of $1/\kappa \approx(10\pm4)\,\mathrm{nm}$ includes uncertainties caused by background variations contaminating the small magnetic signal that is resolved in a limited $Q$ regime.} Comparing panel (a) and (b) in Fig.\,1 we see that the SF data confirm the non polarized magnetic estimates of fluctuations with similar magnitude ($M_y^2$) and length scales  
for common data sets in the higher {$Q$ regime}.
But below  {$Q=0.1~\mathrm{nm}^{-1}$} the smaller SF data are difficult to resolve from dominating NSF data after polarization corrections and do not reveal the signatures of {long-range} order. 
If the $1/Q^4$ upturn in the {$NPH$} difference is real magnetic scattering or an artifact of a nuclear origin or multiple scattering cannot be resolved with SF scattering and needs a different approach.\\   
%
%
%
%
%
%
%

We cannot use the total non spin flip data, NSF, (DD+UU) to reveal the longitudinal magnetic component $M_x^2$ from the angular dependence\cite{Krycka2012}
since the nuclear scattering scattering is dominating the signal. But we can take advantage of the difference response  
%
%
%
%
%
%
``$DIF$'' between the two initial polarization direction (without registering spin flip), the NSF asymmetry or flipper difference from full pol data (DD-UU) and HP data (D-U) that yields an interference term of nuclear and magnetic origin \cite{Krycka2012}.
It signals a weak contribution from a center with a net magnetic component along the $x$-direction $M_x>0$ in the presence of a strong nuclear contribution from the same center.
\begin{equation}
DIF(\theta)= 2NM_x sin^2\theta   
\end{equation}
%
%
%
$DIF(\theta)$ is shown in the inset of Fig.\,4 presenting the two maxima at $DIFV=2NM_x$ according to Eq.\,(4). 
Even in the low {$Q$ regime} this structure can be resolved at low {$T=3$~K} in sufficient high fields {($B\geq50$~mT)}. As shown in the main panel $DIF(Q)=2NM_x(Q)$ can be presented by a $1/Q^4$ term and a small constant following Eq.\,(2) with a large parameter $\kappa$. Since such {$Q$ dependence} is expected for nuclear scattering $N^2$ dominated by large grain boundaries we conclude a similar {$Q$ dependence} for $(M_x)^2$. Potential deviations of the form $1/(K^2+Q^2)^2$ yield magnetic domain sizes larger than {$1/K\approx50$~nm}. 
Fig.\,3(b) presents the estimates of $2NM_x$ collected at different temperatures from the angle dependence ($A$) and the {$Q$ dependence} ($D/Q_m^4$) at low {$Q\approx0.1\,\mathrm{nm}^{-1}$},  $2NM_x\neq0$ for $T<T_c$ while 
$2NM_x\approx0$ for $T\geq T_c$.
This interference term $DIFV$ succeeds to resolve magnetic scattering expected for aligned ferromagnetic magnetic domains that form below $T_c$. The {$1/Q^4$ dependence} of $(M_x)^2$ reveals {long-range} magnetic domains. Simple estimates\cite{Yusuf2000} from neutron depolarization yield domain scales of micrometers at {$T=3$~K}. 
We expect that in a ferromagnet below $T_c$ fluctuations turn into {long-range} order, but in this inhomogeneous compound short-range fluctuations are more dominant than in defect-free, homogeneous systems. On one hand random defects produce distinct short-range correlations that are noticed at $T_c$ and are still present at low temperatures but on the other hand they do not destroy long-range order in this alloy.\\      
%


Collecting SANS data with and without polarization analysis we gained new insight in the inhomogeneous ferromagnetic state of Ni$_{1-x}$V$_x$ with low critical temperatures $T_c$ below 50\,K. In this paper we focus on Ni$_{0.90}$V$_{0.10}$. We found clear evidence of magnetic fluctuations in the larger {$Q$ regime} from spin flip (SF) contrast. From the magnetic fluctuations at $T_c$ a fraction of 10\% remains at the lowest temperature of {$T=3$~K} with similar correlation lengths of about {10\,nm}. In addition, the non spin flip (NSF) asymmetry (from full pol and half pol data) reveals large scale aligned magnetic domains in the lower {$Q$ regime} at low temperatures {$T=3$~K} below $T_c$. 
Although these random defects cause short-range magnetic fluctuating clusters, long-range order still develops in this alloys. 
Similar features can be observed in Ni$_{1-x}$V$_x$ samples with $x=0.11$ with smaller {$T_c\approx7$~K}. The challenge to resolve the smaller magnetic contribution from the overwhelming nuclear background is increased, but magnetic fluctuations remain and indication of aligned domains are present for low temperatures below $T_c$.
More details will be presented elsewhere\cite{Shiva}.
We demonstrated that PASANS is a helpful method to clarify signatures of random dilution in alloys 
presenting magnetic correlations that persist in a wide range of length scales at low temperatures.\\ 
 %

We thank J. Kryzwon, T. Dax, S. Watson and T. Hassan for their support with NG7SANS, cryogenics and $^3$He cell spin filters preparation at NIST.
Support for  usage of the {$^3$He} spin polarizer on the NG7 SANS instrument was provided by the Center for High Resolution Neutron Scattering, a partnership between the National Institute of Standards and Technology and the National Science Foundation under Agreement No. DMR-1508249.
This research is funded in part by a QuantEmX grant from ICAM and the Gordon and Betty Moore Foundation through Grant GBMF5305 to Hector D. Rosales.
We thank Lisa DeBeer-Schmitt for her support at GPSANS, ORNL. 
A portion of this research used resources at the High Flux
Isotope Reactor, which are DOE Office of Science User
Facilities operated by Oak Ridge National Laboratory.
%

%

\end{document}